\documentclass[aps,preprint,nofootinbib]{revtex4}

\newcommand{\be}{\begin{eqnarray}}
\newcommand{\ee}{\end{eqnarray}}
\newcommand{\bitem}{\begin{itemize}}
\newcommand{\eitem}{\end{itemize}}
\newcommand{\non}{\nonumber\\}
\newcommand{\inline}[1]{\noalign{\hbox{#1}}}
\newcommand{\hatH}{{\hat{H}}}
\newcommand{\hatU}{{\hat{U}}}
\newcommand{\bra}[1]{\langle #1 |}
\newcommand{\ket}[1]{| #1 \rangle}
\newcommand{\ave}[1]{\langle #1 \rangle}
\newcommand{\hatrho}{{\hat{\rho}}}
\newcommand{\tildephi}{{\tilde{\phi}}}
\newcommand{\calZ}{{\cal Z}}
\newcommand{\calD}{{\cal D}}
\newcommand{\calJ}{{\cal J}}
\newcommand{\hatphi}{{\hat{\phi}}}
\newcommand{\bfp}{{\bf p}}
\newcommand{\bfk}{{\bf k}}
\newcommand{\bfx}{{\bf x}}
\newcommand{\tw}{\textwidth}

\usepackage{epsf}

\begin{document}

\preprint{McGill-Nuc}

\title{The Boltzmann Equation in Classical and Quantum Field Theory}

\author{Sangyong Jeon$^{\dagger, *}$}

\affiliation{$^\dagger$Physics Department, McGill University, 3600
University Street, Montreal, QC H3A-2T8, Canada}%
\affiliation{$^*$RIKEN-BNL Research Center, Brookhaven National
Laboratory, Upton, NY 11973}

\begin{abstract}
 
 Improving upon the previous treatment by Mueller and Son,
 we derive the Boltzmann equation that results from a classical scalar field
 theory.  This is obtained by starting from
 the corresponding quantum field theory and taking the classical limit
 with particular emphasis on the path integral and perturbation theory.
 A previously overlooked Van-Vleck determinant is shown to control the 
 tadpole type of self-energy that can still appear in the classical
 perturbation theory. 
 Further comments on the validity of the approximations and possible
 applications are also given.

\end{abstract} 

\maketitle

\section{Introduction}

There are many situations in nature where classical field theory
is the most direct and efficient way of describing the system
under study. This happens when the occupation number becomes
large enough that the quantum commutator becomes irrelevant.
Such situations include most of macroscopic electro-magnetic
systems and the dense systems created in colliding heavy ions. 
In studying heavy ion collisions, both the particle degrees of freedom
and the classical field degrees of freedom becomes relevant:
The final state of a collision consists mostly of the
particle degrees of freedom
whereas the initial state of two approaching nuclei can be
efficiently described by a classical non-Abelian gauge field 
\cite{MV,Kovner:1995ts,Kovner:1995ja}.
A natural framework to encompass both elements is
the coupled system of classical field equation and the Boltzmann
type of kinetic equation with the mean field\cite{Blaizot:1999xk}.

An important problem in this framework is the conversion of classical field
degrees of freedom into particle degrees of freedom and their subsequent
thermalization. In this respect, the equivalence of the 
classical field theory and the
Boltzmann equation proposed by Mueller and Son\cite{MS} is significant
since it has a potential of providing a consistent
framework for the thermalization in heavy ion collisions.
The current paper improves on the work of Mueller and Son in the 
following aspects.

First, going from quantum to
classical many body theory almost always involves the Wigner
function\,\cite{Calzetta:1986cq,Mrowczynski:1994nf,Blaizot:1999xk}.
In Ref.\cite{MS}, this point was overlooked.
In this work, we show that the Wigner functional does appear in the
formulation due to the fact that the density operator is in general
non-diagonal.

Second, the Feynman rules used in Ref.\cite{MS} for the classical field are
a mixture of quantum and classical ones: The vertex rule is
classical but the propagators are quantum mechanical.
We show that consistent formulation only involves classical vertices and
propagators.

Third, formulation of classical path integral involves a functional
determinant related to the Jacobi field. 
Therefore, perturbation
theory based on classical path integral involves ghost fermions
\cite{Deotto:2002hy,Deotto:2002hz}.
It turned out that the classical perturbation theory as formulated in
Ref.\cite{MS} still contains a quantum effect in the form of 
the tadpole self-energy. 
The ghost contribution is shown to compensate this remaining quantum effect,
thus making the theory consistently classical.

Despite these gaps in the formulation,
the final result in Ref.\cite{MS} is essentially correct in the
$f \gg 1$ limit.
The goal of this paper is to provide a more consistent derivation of
the kinetic equation corresponding to the classical scalar field 
starting from the quantum scalar field.
We take particular care in treating the path integral and the perturbation
theory in a consistent manner.
We also comment on the validity of this classical method and
possible use in the context of understanding the thermalization in heavy ion
collisions.

 \section{Classical Path Integral from Quantum Path Integral}

To be specific,
consider a real scalar field theory defined by the following Hamiltonian 
 \be
 H = 
 \int d^3 x
 \left( {\pi^2\over 2} + {(\nabla \phi)^2\over 2} + V(\phi) \right)
 \ee
 The equations of motion are
 \be
 \dot{\phi} &=& {\delta H \over \delta \pi} = \pi
 \label{eq:dotphi}
 \\
 \dot{\pi} &=& -{\delta H\over \delta \phi} \equiv E(\phi)
 \label{eq:dotpi}
 \ee
 where we defined $E(\phi) = \nabla^2\phi - V'(\phi)$ for notational
 convenience.

In many publications since 1980's E.~Gozzi and his collaborators have
been extensively 
studying the properties of classical path integrals 
(See Refs.\cite{Deotto:2002hy,Deotto:2002hz} and references therein.).
The starting point in these studies is the following transition probability 
between two points in the phase space
\be
P[\varphi_f,\pi_f^\varphi;t_f| \varphi_i,\pi^\varphi_i;t_i]
=
\delta[\varphi_f - \varphi_c(t_f| \varphi_i,\pi^\varphi_i)] \, 
\delta[\pi^\varphi_f - \pi^\varphi_c(t_f| \varphi_i,\pi^\varphi_i)] \, 
\ee
Here $\varphi$ and $\pi^\varphi$ are the generalized coordinate and momentum
and $\varphi_c$ and $\pi^\varphi_c$ are the solutions
of Eqs.~(\ref{eq:dotphi}) and (\ref{eq:dotpi})
with the
boundary conditions $\varphi_c(t_i|\varphi_i,\pi^\varphi_i) = \varphi_i$ and 
$\pi^\varphi_c(t_i|\varphi_i,\pi^\varphi_i) = \pi^\varphi_i$.
The evolution of the 
initial phase space density 
$\rho_{\rm cl}[\varphi, \pi^\varphi; t_i]$ is given by
\be
\rho_{\rm cl}[\varphi, \pi^{\varphi}; t]
=
\int [d\varphi_i][d\pi^\varphi_i]\,
P[\varphi,\pi^\varphi;t| \varphi_i,\pi^\varphi_i;t_i]\,
\rho_{\rm cl}[\varphi_i, \pi^\varphi_i; t_i]
\label{eq:rhoclt}
\ee
In this paper we use the square bracket notation $\int [d\phi]$ to indicate
a functional integral at a fixed time and the curly $D$ notation
$\int {\cal D}\phi = \int \prod_{i=1}^N [d\phi_i]$ 
to indicate a functional integral over space and time.
We also implicitly absorb any normalization constants into the definition of
$[d\phi]$.

To turn Eq.(\ref{eq:rhoclt}) into a path integral,
we use the fact that transition probabilities satisfy 
\be
\int dy\, P[z|y]P[y|x] = P[z|x]
\ee
 Dividing the time between the final time $t_f = t_{N+1}$
 and the initial time $t_i = t_0$ 
 into $N{+}1$ equal intervals we obtain
 \be
 P[\phi_{N+1}, \pi_{N+1}| \phi_0, \pi_0]
 =  
 \int 
 \prod_{i=1}^{N} 
 [d\phi_i][d\pi_i]\, 
 \prod_{j=0}^{N} 
 P[\phi_{j+1}, \pi_{j+1}| \phi_j, \pi_j]
 \ee
 where we have suppressed the time arguments in $P$.

 For sufficiently small $\Delta t = (t_f-t_i)/(N{+}1)$, 
 we should be able to
 solve the classical equations of motion approximately.
 The simplest finite difference method is the Euler method
 based on the following approximation of the time derivative
 \be
 \dot{f}(t_k) \approx {f(t_k) - f(t_{k-1})\over \Delta t} + O(\Delta t^2)
 \ \ \ \hbox{(Backward Euler)}
 \\
 \inline{or}
 \dot{f}(t_k) \approx {f(t_{k+1}) - f(t_{k})\over \Delta t} + O(\Delta t^2)
 \ \ \ \hbox{(Forward Euler)}
 \ee
 We use here the Backward Euler method for $\varphi$
 \be 
 \varphi_{k+1}
 & = &
 \varphi_k + \pi^\varphi_{k+1}\,\Delta t + O(\Delta t^2)
 \label{eq:varphi_kp1}
 \ee
 and the Forward Euler method for $\pi^\varphi$ 
 \be
 \pi^\varphi_{k+1}
 & = &
 \pi^\varphi_k + E(\varphi_{k})\, \Delta t + O(\Delta t^2) 
 \label{eq:varpi_kp1}
 \ee
 so that $\varphi_{k+1}$ can be expressed solely in terms of $\varphi_k$
 and $\pi^\varphi_k$.  This is, of course,
 not the only choice. Different 
 discretization method in general leads to different form of the discretized 
 path integral although they are all equivalent in the $\Delta t\to 0$ limit.

 For small enough $\Delta t$ then, 
 \be
 P[\varphi_{j+1}, \pi^\varphi_{j+1}|\varphi_j, \pi^\varphi_j]
 =
 \delta[\varphi_{j+1} - \varphi_j - \pi^\varphi_{j+1}\Delta t]
 \delta[\pi^\varphi_{j+1} - \pi^\varphi_j - E(\varphi_{j})\Delta t]
 \ee
 With this form, it is easy to check that the preservation of probability
 \be
 \int [d\varphi_{N+1}][d\pi^\varphi_{N+1}]\,
 \rho_{\rm cl}[\varphi_{N+1}, \pi^\varphi_{N+1}; t_{N+1}]
 =
 \int [d\varphi_{0}][d\pi^\varphi_{0}]\,
 \rho_{\rm cl}[\varphi_{0}, \pi^\varphi_{0}; t_{0}]
 \ee
 is trivially satisfied.
 Using dummy variables $\chi$ and $\pi^\chi$ to express the
 $\delta$-functionals, we finally get
 the classical path integral for the evolution of the density functional
 \be
 \rho_{\rm cl}[\varphi_{N+1}, \pi^\varphi_{N+1}; t_{N+1}]
 & = &
 \int \prod_{j=0}^N [d\varphi_j] 
 \prod_{k=1}^{N+1} [d\pi^\chi_k]
 \exp
 \left(
 i\sum_{j=0}^{N}
 \pi^\chi_{j+1} 
 \left(\varphi_{j+1} - \varphi_j - \pi^\varphi_{j+1}\Delta t)
 \right)
 \right)
 \non
 & & {} \times
 \int \prod_{l=0}^N [d\pi^\varphi_l]\, 
 \prod_{k=0}^{N} [d\chi_k]\, 
 \exp
 \left(
 -i\sum_{k=0}^{N}
 \chi_k
 \left(
 \pi^\varphi_{k+1} - \pi^\varphi_k - E(\varphi_{k})\Delta t
 \right)
 \right)
 \non
 & & {} \times
 \rho_{\rm cl}[\varphi_0, \pi^\varphi_0; t_0]
 \label{eq:rho_cl_f}
 \ee

 How closely can we reproduce the classical path integral 
 Eq.(\ref{eq:rho_cl_f}) from the corresponding
 quantum path integral? 
 To answer this question,
 we need to go back to the basics of deriving a quantum mechanical
 path integral.
 Given a Hamiltonian operator
 $\hatH$, we would like to rewrite the matrix element 
 of the evolution operator, $\bra{\phi_{f}} \hatU(t_f, t_i) \ket{\phi_i}$,
 as a path integral.
 Dividing the time interval into many small segments and using the fact that
 $\hatU(t, t')\hatU(t',t'') = \hatU(t,t'')$, we can write
 \be
 \bra{\phi_{N+1}} \hatU(t_{N+1}, t_0) \ket{\phi_0}
 & = &
 \bra{\phi_{N+1}} 
 \hatU(t_{N+1}, t_N) 
 \hatU(t_N, t_{N-1}) 
 \cdots
 \hatU(t_1, t_0) 
 \ket{\phi_0}
 \ee
 There are many ways of inserting the resolutions of identity to turn this
 expression into a path integral just as there are many ways of discretizing
 the classical equation of motion.  
 The prescription which most closely resembles 
 Eqs.(\ref{eq:varphi_kp1}) and (\ref{eq:varpi_kp1})
 turned out to be inserting
 \be
 1 = \int [d\phi_k][d\pi_k]
 \ket{\phi_k}\ave{\phi_k|\pi_k}\bra{\pi_k}
 \ee
 between $\hatU(t_{k+1},t_k)$ and $\hatU(t_k, t_{k-1})$ for all
 $1\le k \le N$ and inserting
 \be
 1 = \int [d\pi_{N+1}]
 \ket{\pi_{N+1}}\bra{\pi_{N+1}}
 \ee
 between $\bra{\phi_{N+1}}$ and $\hatU(t_{N+1}, t_N)$.
 In this way, we get
 \be
 \bra{\phi_{N+1}} \hatU(t_{N+1}, t_0) \ket{\phi_0}
 & = &
 \int \prod_{k=1}^N [d\phi_k]\, 
 \int \prod_{k=1}^{N+1} [d\pi_k]\, 
 \non & & {} \times
 \exp\left(
 i\sum_{k=1}^{N+1}\pi_k (\phi_k - \phi_{k-1})
 -i\sum_{k=1}^{N+1}  H(\pi_k, \phi_{k-1}) h
 \right)
 \ee

 The time evolution of a density operator needs two such path integrals
 since the matrix element
 \be
 \bra{\phi_f}\hatrho(t_f) \ket{\tildephi_f}
 & = &
 \bra{\phi_f}\hatU(t_f,t_i)\hatrho(t_i) \hatU(t_i, t_f) \ket{\tildephi_f}
 \non
 & = &
 \int [d\phi_i][d\tildephi_i]\,
 \bra{\phi_f}\hatU(t_f,t_i)\ket{\phi_i}
 \bra{\phi_i} \hatrho(t_i) \ket{\tildephi_i}
 \bra{\tildephi_i}\hatU(t_i, t_f) \ket{\tildephi_f}
 \ee
 involves two matrix elements of the evolution operator $\hatU$.
 To bring out classical features more clearly, define
 \be
 \varphi &=& (\phi + \tildephi)/2
 \\
 \chi &=& \phi - \tildephi 
 \ee
 where $\phi$ belongs to the path integral representation of
 $\bra{\phi_f}\hatU(t_f,t_i)\ket{\phi_i}$ and
 $\tilde{\phi}$ belongs to the path integral representation of 
 $\bra{\tildephi_i}\hatU(t_i, t_f) \ket{\tildephi_f}$.

 Introducing the Wigner functional $\rho_W(\varphi, \pi^\varphi)$ as
 \be
 \bra{\varphi + \chi/2} \hatrho
 \ket{\varphi - \chi/2} 
 =
 \int [d\pi^\varphi]\, \rho_W[\varphi, \pi^\varphi] \, 
 e^{i\chi \pi^\varphi}
 \ee
 we get
 \be
 \rho_W[\varphi_{N+1}, \pi^\varphi_{N+1}]
 & = & 
 \int [d\varphi_0][d\chi_0]\,
 \rho_W[\varphi_0, \pi^\varphi_0]
 \non && {} \times
 \int \prod_{k=1}^N [d\varphi_k]\, 
 \int \prod_{k=1}^{N} [d\chi_k]\, 
 \int \prod_{k=0}^{N} [d\pi^\varphi_k]\, 
 \int \prod_{k=1}^{N+1} [d\pi^\chi_k]\, 
 \non & & {} \times
 \exp\left(
 i\sum_{k=0}^{N}
 \pi^\chi_{k+1} (\varphi_{k+1} - \varphi_{k} - \pi^\varphi_{k+1} \Delta t)
 \right)
 \non & & {} \times
 \exp\left(
 -i\sum_{k=0}^{N}\chi_k 
 \left(\pi^\varphi_{k+1} - \pi^\varphi_{k}
 - E(\varphi_k) \Delta t 
 \right)
 + O(\chi^3)
 \right)
 \label{eq:rho_W_fin}
 \ee
 Note that without the Wigner transformation of the initial and the final 
 distributions, the exponents cannot be arranged in the above 
 finite-difference form that includes the end points.

 The difference between Eq.(\ref{eq:rho_W_fin}) and the classical
 expression Eq.(\ref{eq:rho_cl_f}) is the $O(\chi^3)$ term inside the
 exponential. 
 (There are no terms even in $\chi$ in this expression because
 the exponent must change sign under the exchange $\phi \leftrightarrow
 \tildephi$ or equivalently, $\chi\to -\chi$.)
 If we drop the $O(\chi^3)$ terms, then we have exactly the same
 {\em evolution kernel} as the classical theory.
 However, even in this case, this does not mean that $\rho_W$ will be truly
 classical.  If the initial Wigner functional contains quantum information,
 then $\rho_W$ at any later times will still contain it.

 In perturbation theory, dropping $O(\chi^3)$ terms means ignoring any 
 Feynman diagrams that may contain a loop or loops made up of the {\em
 off-shell} propagators.  
 The validity of such operation will be shortly discussed in
 the next section where we discuss the Feynman rules.

 To summarize this section, the omission of the terms non-linear in
 $\chi$ results in the classical {\em evolution} of the initial distribution.
 The crucial difference between our analysis here and that of Ref.\cite{MS}
 is the appearance of the Wigner functional
 which depends both on the coordinate
 $(\varphi)$ and the momentum $(\pi^\varphi)$.  
 This is important in the perturbation theory
 since the propagators strongly depend on the initial distribution.
 We therefore discuss the propagators and the Feynman rules next.

\section{Propagators and Feynman Rules}
\label{sec:props}

 Having obtained the {\em classical} path integral from a quantum one, 
 we can now derive Feynman rules for the classical perturbation
 theory following the same procedure as in the quantum case. 
 To do so, consider the generating functional with the source terms 
 $J_\varphi \varphi - J_\chi\chi$ and the restrictions that the source
 terms vanish at the end points $t_i$ and $t_f$.
 Carrying out $\pi^\chi$ integrals and $\chi_0$ integral results in 
 $$ 
 \delta\left[
 {\varphi_{1}-\varphi_0\over h} - \pi^\varphi_{0}
 - E(\varphi_0) h 
 \right]
 \prod_{k=0}^N
 \delta\left[
 \varphi_{k+1} - \varphi_{k} - \pi^\varphi_{k+1} h
 \right]
 $$ 
 We can then carry out all $\pi^\varphi_{k}$ integrals to get
 \be
 \calZ[J_\varphi, J_\chi]
 & = &
 \int \prod_{k=0}^{N+1} [d\varphi_k]\, 
 \int \prod_{k=1}^{N} [d\chi_k]\, 
 \rho_W\left[\varphi_0, (\varphi_{1}-\varphi_0)/h - E(\varphi_0) h \right]
 \non && {} \times
 \exp\left(
 -i\sum_{k=1}^{N}\chi_k 
 \left( {\varphi_{k+1} - 2\varphi_{k}+ \varphi_{k-1}\over h}
 - E(\varphi_k) h + J^\chi_k h
 \right)
 +
 i\sum_{k=1}^{N} J^\varphi_k \varphi_k h
 \right)
 \non
 & = &
 \int {\calD \varphi}\,{\cal D\chi}\, \rho_W[\varphi_i, \dot{\varphi}_i]\,
 \non & & {} \times
 \exp\left(
 i\int \, 
 \chi\left(-\partial^2 \varphi - m^2 \varphi - {\lambda\over 3!} \varphi^3
 -J_\chi 
 \right)
 + i\int J_\varphi \varphi
 \right)
 \ee
 where we switched to the continuum notation for simplicity.

 It is tempting now to develop a perturbation theory from this expression. 
 But we are not done yet. Note that the initial distribution 
 depends on the value of the field and its time derivative.  But 
 the integration is over the field values only.  
 To treat $\rho_W[\varphi_i, \dot{\varphi}_i]$ as a proper 
 weight we need to transform the measure to include integration over
 $\dot{\varphi}_i$. 
 In classical field theory,
 this can be most easily done by noting that to solve 
 a second order differential equation requires
 specifying either the values of the field and its first time
 derivative at the initial time
 or the values of the field at the initial and the final time.
 Therefore, barring caustic points, there is a
 one-to-one correspondence between the value of $\varphi$
 at $t_f$ and the
 value of $\dot{\varphi}$ at $t_i$.  The Jacobian of this transformation is
 \be
 \calJ =
 \left| {\delta \varphi_f \over \delta \dot{\varphi}_i} \right|
 =
 \left|
 {\delta^2 S\over \delta \varphi_i \delta \varphi_f} 
 \right|^{-1}
 \ee
 where $S$ is the action integral. 
 This Jacobian is related to the Van-Vleck determinant
 \be
 \calJ[\varphi]
 =
 \left| 
 {\rm Det}
 (\partial^2 + m^2 + (\lambda/2)\varphi^2) 
 \right|
 \label{eq:van-vleck}
 \ee

 Changing the variable $\varphi_f$ to $\dot{\varphi}_i$ then gives 
 \be
 {\cal Z}[J_\varphi, J_\chi]
 &=& 
 \int [d\pi_i] [d\varphi_i] \rho_W[\varphi_i, \pi_i] \,
 \int_{\varphi_i,\pi_i} {\cal D}\varphi {\cal D}\chi\,
 \non & & {} \times
 \calJ[\varphi]\,
 \exp\Bigg[
 i\int \chi\left(- \partial^2 \varphi - m^2 \varphi
 - {\lambda\over 3!} \varphi^3 - J_\chi\right)
 + i\int J_\varphi \varphi 
 \Bigg] 
 \label{eq:final_calZ}
 \ee
 where we have renamed $\dot{\varphi}_i = \pi_i$ and there is no 
 $[d\varphi_f][d\chi_f]$ integral. 
 One can easily check that in the limit $J_\varphi\to 0$, 
 ${\cal Z} \to 1$.
 This is as it should be since 
 $\calZ$ reduces to a trace of a density operator in this limit. 
 
 Note that without the Jacobian $\calJ$, the $J_\varphi\to 0$ limit will not
 in general yield 1. This is because
 performing $\chi$ integral produces 
 $\delta[ (\partial^2\varphi + m^2\varphi + (\lambda/3!)\varphi^3 ) + J_\chi ]$
 which is then integrated over $\varphi$ to yield $1/\calJ$.
 The exact value of the Jacobian, however, depends on how one discretizes the
 equation of motion.  This may sounds peculiar, but this phenomenon is not
 new. In stochastic dynamics, it is well known that this kind of functional
 Jacobian depends on the discretization prescription\cite{drozdov}.
 Furthermore, the Jacobian 
 can be made constant if one choose a particular
 prescription. This was also noticed in Ref.\cite{Gozzi:2000sf}.
 Later in this section,
 we will comment more on the role of the determinant.
 For now we simply set it to a constant and ignore it.
 As shown shortly, this amounts to the following diagrammatic rule:
 \bitem
 \item[(i)] Omit diagrams containing tadpoles.
 \eitem

 To get the rest of the Feynman rules, 
 let us examine the two main ingredients of perturbation theory:
 Propagators and interaction vertices.  
 The forms of the interaction vertices are fully
 determined by the interaction Lagrangian. 
 In the case of the classical $\lambda\phi^4$ theory,
 the corresponding Feynman rule is:

 \bitem
 \item[(ii)] Assign $-i\lambda$ to 
 each $\lambda\varphi^3\chi$ vertex. 
 \eitem
 To determine propagators, one must
 specify what the {\em un-perturbed state} is.
 In the quantum theory, this is the perturbative vacuum which
 can fluctuate into particle-antiparticle pairs. Therefore,
 the expectation value $\bra{0}\hatphi(x)\hatphi(y)\ket{0}$ is non-zero even
 if $\bra{0}\hatphi(x) \ket{0} = 0$.
 On the other hand, in the classical vacuum, both of these quantities are
 zero because the classical vacuum cannot fluctuate. It is
 literally a state where nothing exists.

 Propagators are determined by the free field limit.
 Following the derivation in the last section, it is easy to see that without
 the self-interaction, the generating function quickly reduces to
\be
 {\cal Z}[J_\varphi,J_\chi]
 & = &
 \int [d\pi_i] [d\varphi_i]
 \rho_W[\varphi_i, \pi_i] \,
 \non & & {} \times
 \int_{\varphi_i,\pi_i} {\cal D}\varphi 
 \calJ[\varphi]\,
 \delta\left[- \partial^2 \varphi - m^2 \varphi + J_\chi\right]
 \exp\Bigg[
 i\int J_\varphi \varphi 
 \Bigg] 
\ee
regardless of whether it's the quantum case or the classical case.
Yet for both cases, the combination
$\varphi = (\phi + \tildephi)/2$ satisfies the classical equation of motion
whose solution is given by
\be
\varphi = 
 \left(
 \varphi_i \cos(E_k (t -t_i))
 +
 \pi_i {\sin(E_k (t-t_i))\over E_k}
 \right)
 +
 \int_{t_i}^t dt'\, G_{\rm ret}(t-t') J_\chi(t')
 \label{eq:varphi_cl}
\ee
The fact that $\varphi$ satisfies the classical equation of motion
even in the fully quantum case is somewhat unexpected.
However, it shows explicitly that the form of the initial distribution
functional $\rho_W$ plays an essential role in distinguishing
the quantum and the classical cases.

 Let's consider the purely classical case first. 
 In equilibrium, the density functional is given by the classical Boltzmann
 factor
 \be
 \rho_{\rm cl}[\varphi_i, \pi_i]
 = {e^{-\beta H_{\rm cl}}\over Z_{\rm cl}}
 =
 {1\over Z_{\rm cl}}
 \exp\left(-\beta \sum_k (\pi_i(k)^2 + E_k^2 \varphi_i(k)^2)/2 \right)
 \label{eq:cl_Boltz}
 \ee
 where $\pi_i(k)$ and $\varphi_i(k)$ are the Fourier components of $\pi_i$ and
 $\varphi_i$ respectively.
 Using the above classical solution and
 carrying out the $\varphi_i$ and $\pi_i$ integral then yield
 \be
 \calZ_{\rm cl}[J_\varphi, J_\chi]
 &=&
 \exp\Bigg[
 -\sum_k {T\over 2E_k^2}
 \int_{t_i}^{t_f} dt
 \int_{t_i}^{t_f} dt'\,
 J_\varphi(t) \cos(E_k(t-t')) J_\varphi(t')
 +
 i\int J_\varphi G_{\rm ret} J_\chi
 \Bigg]
 \non
 \ee
 where we have again 
 suppressed the momentum indices for $J_\varphi$ and $J_\chi$.
 The resulting Feynman rules are: 
 
 (iii) In equilibrium, the propagators are
 \be
 G_{\rm cl}^{\varphi\varphi}(p) & = & (T/E_p) 2\pi\,\delta(p^2 - m^2) 
 \\
 G_{\rm cl}^{\varphi\chi}(p) & = &  {i\over p^2 - m^2 + i\epsilon p_0}
 \\
 G_{\rm cl}^{\chi\varphi}(p) & = &  {i\over p^2 - m^2 - i\epsilon p_0}
 \\
 G_{\rm cl}^{\chi\chi}(p) & = & 0
 \ee
 The same set of rules were also used in  
 Refs.\cite{Aarts:1997kp,Aarts:1999wj,Berges:2002wf,Berges:2004yj}
 as the classical limit of the quantum rules.

 The $\varphi\chi$ propagator is the response function which controls
 generation of classical field by a source. Hence it must be there be it
 classical or quantum ({\it c.f.}~Eq.(\ref{eq:varphi_cl})).
 However, in the $T\to 0$ limit, $\varphi\varphi$ propagator vanishes due to
 the fact that classical vacuum does not fluctuate.
 Note also that the
 classical equilibrium density function is $T/E_k$ just as in the case of
 electromagnetic waves in a cavity. 
 Therefore, if one is to extend this formalism to an out-of-equilibrium
 situation, one should use 
 
 (iii${}'$) In non-equilibrium, 
 \be
 G_{\rm cl}^{\varphi\varphi}(p, X) & = & 
 f_{\rm cl}(p, X)\, 2\pi\,\delta(p^2 - m^2) 
 \ee
 where $f_{\rm cl}(p, X)$ is the density function to be determined and
 $G_{\rm cl}^{\varphi\varphi}(p, X)$ is the 
 Wigner transformed 2 point function
 \be
 G_W(p, X) = \int d^4 r\, e^{i r^\mu p_\mu}\, G(x, y)
 \label{eq:G_Wigner}
 \ee
 with $r = x-y$ and $X = (x+y)/2$.

 The appearance of the on-shell $\delta$-function in 
 $G_{\rm cl}^{\varphi\varphi}$ propagator deserves deeper consideration.
 Suppose that the initial $\rho[\varphi_i, \chi_i]$ is actually independent 
 of $\chi_i$. That is, $\hat{\rho}$ is diagonal in $\phi$. 
 In that case, the Wigner functional contains a $\delta$
 functional $\delta[\pi_i]$.  For an illustration, suppose 
 \be
 \rho_W'[\varphi_i,\pi_i]
 = {1\over Z'} \exp\left(-\beta\sum_k E_k^2 \varphi_i(k)^2/2\right) 
 \delta[\pi_i]
 \ee
 In that case, carrying out $\varphi_i$ and $\pi^\varphi_i$ 
 integrals results in
 \be
 \calZ'[J_\varphi, J_\chi]
 &=&
 \exp\Bigg[
 -\sum_k {T\over 2E_k^2}
 \left( \int_{t_i}^{t_f} dt
 J_\varphi(t) \cos(E_k(t-t_i))
 \right)^2
 +
 i\int J_\varphi G_{\rm ret} J_\chi
 \Bigg]
 \non
 \ee
 Since there is no compensating sine term, the $\varphi\varphi$ Green
 function no longer exhibits time-translation invariance.  
 The $\varphi\varphi$ correlator is given by
 \be
 G^{\varphi\varphi}(t, t')
 =
 {T\over E_k^2} \cos(E_k(t-t_i)) \cos(E_k(t'-t_i))
 \ee
 Since this is not a function of $t-t'$, this form of the propagator
 {\em does not} conserve energy at each vertex. 
 The usual momentum space Feynman diagram techniques will not work.

 Our ability to use Feynman rules with the usual conservation
 4-momentum at each vertex depends on whether we get a function
 only of $t-t'$ for the $\varphi\varphi$ correlator.
 In equilibrium, the propagators depend only on the difference of the two
 coordinates.  Hence, the total 4-momentum is conserved at each vertex.
 In non-equilibrium situations, the propagators depend both on 
 the difference $r = x-y$ and the sum $X = (x+y)/2$.
 Wigner transformation assigns `momentum' to the difference
 as done in Eq.(\ref{eq:G_Wigner}).
 However, unless the $X$ dependence of $G_W(p, X)$ is slow, 
 the momenta that enter each vertex are not conserved. For the derivation of
 the Boltzmann equation, this slow dependence is therefore essential.

 The above Feynman rules are used to calculate the perturbative corrections
 in the classical field theory.  Just as loops appear 
 in the quantum Feynman diagrams, 
 loops appear in the classical Feynman diagrams as well~\cite{MS}.
 This looks like a violation of the theorem that states
 the classical field theory corresponds to the sum of all tree
 diagrams. However, this theorem is derived in the context of the vacuum
 theory.  It does not necessarily apply to the in-medium case. 
 Furthermore, the theorem is derived considering only the un-cut diagrams
 where as the in-medium theory necessarily involves cut diagrams 
 (that include $\varphi\varphi$ propagators).
 Nevertheless, if this theorem is not to be violated, we must at least show
 that the vacuum loops do not appear in our classical theory.

 That this is indeed the case can be shown as follows.  
 A vacuum loop in a given diagram must not contain the density function
 $f_{\rm cl}$. Since $f_{\rm cl}$ appears only in the $\varphi\varphi$
 propagator, any potential vacuum loop diagram must consist entirely of
 $\chi\varphi$ and $\varphi\chi$ propagators.
 Now with only the
 $\lambda\chi\varphi^3$ vertex, $\chi$ is conserved in the sense that it
 cannot split into multiple $\chi$'s. Hence one can follow $\chi$ as if
 it is a fermion.  What flows along this line is the time since the
 $\varphi\chi$ propagator is a retarded one and the
 $\chi\varphi$ propagator is an advanced
 one.  Therefore, a loop made of $\varphi\chi$ and/or $\chi\varphi$
 propagators implies existence of a time-loop which is impossible.  It
 follows that there is no vacuum loop contribution in the classical
 theory if the loop involves more than a single propagator.

\begin{figure}[thb]
\epsfysize=3cm
\begin{center}
\epsfbox{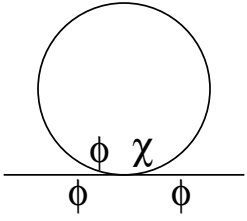}
\end{center}
\caption{Tadpole diagram in the classical perturbation theory.}
\label{fig:tadpole}
\end{figure}
 
 The only remaining possibility is a single loop formed by a $\varphi\chi$
 propagator as shown in Fig.\ref{fig:tadpole} which corresponds to
 \be
 \Sigma_{\rm 1-loop} = {\lambda\over 2} G_{\rm ret}(t=0)
 \label{eq:sigma_1}
 \ee
 The value of this expression is actually ambiguous as it
 involves $\theta(0)$. 
 This is where the Van-Vleck determinant Eq.(\ref{eq:van-vleck}) plays its
 role.
 To determine the role of $\calJ$ in perturbation theory, we may 
 exponentiate the determinant using Fadeev-Popov type of ghost fields $c$
 and $\bar{c}$
 \be
 {\rm Det}(\partial^2 + m^2 + (\lambda/2)\varphi^2)
 \to
 \int \calD c\, \calD \bar{c}\, 
 \exp\left(
 i\int \bar{c} (\partial^2 + m^2 + (\lambda/2)\varphi^2) c
 \right)
 \ee
 Since there is no in-medium part for the ghost field, the corresponding
 propagators should be
 \be
 G_{\rm Q}^{c\bar{c}} & = &  {i\over p^2 - m^2 + i\epsilon p_0}
 \\
 G_{\rm Q}^{\bar{c}c} & = &  {i\over p^2 - m^2 - i\epsilon p_0}
 \ee
 with the rule that a ghost-loop gets an additional factor of
 ($-1$). 

 First of all, ghost propagators must form
 a loop since no external line can be a ghost and there can be no ghost
 excitations in the medium.  
 However, since the ghost propagator is a
 retarded one, a ghost loop made of 2 or more ghost lines must vanish.
 A loop made up of a single ghost line  is then the only possible non-zero
 diagram
 \be
 \Sigma_{\rm 1-loop}^{\rm ghost} = -{\lambda\over 2} G_{\rm ret}(t=0)
 \ee
 This exactly cancels the tadpole contribution in Eq.(\ref{eq:sigma_1}).
 Thus the ghost saves
 the classicality of the perturbation theory
 by removing the ambiguity of defining $\theta(0)$.
 If we define $\theta(0) = 0$, then the ghost field contribution disappears
 and the Van Vleck determinant can be set to a constant.
 Again, as mentioned earlier, this is a known phenomenon.  In the study of
 stochastic dynamics, it is well known that how one discretizes can change
 the form of Van Vleck determinant even making it a constant\cite{drozdov}.
 In Ref.\cite{MS}, since all loops that are formed with $\varphi\chi$
 propagators and/or $\chi\varphi$ propagators
 are set to zero, one might say that 
 $\theta(0)$ was implicitly defined to be zero in that paper.

 To determine the validity of the above classical perturbation theory as a
 approximation of the underlying quantum theory, 
 let us now consider the quantum case.  The form of the free field 
 Wigner function 
 can be easily obtained from the ground state wavefunction of the simple
 harmonic oscillator in the Euclidean space (For instance, see
 Ref.\cite{FH}.).
 \be
 \rho_{\rm Q}[\varphi_i, \pi_i]
 =
 {1\over Z_Q}
 \exp
 \left(
 -\sum_k {\tanh(E_k \beta/2)\over E_k}
 \left(\pi_i(k)^2 + E_k^2 \varphi_i(k)^2\right)
 \right)
 \ee
 In the small $\beta$ or large $T = 1/\beta$ limit, this goes over to the
 classical case.
 Again using the classical solution and 
 carrying out the $\varphi_i$ and $\pi_i$ integral yield
 \be
 \calZ_{Q}[J_\varphi, J_\chi]
=
 \exp\Bigg[
 -\sum_k {\coth(E_k\beta/2)\over 4E_k}
 \int_{t_i}^{t_f} dt
 \int_{t_i}^{t_f} dt'\,
 J_\varphi(t) \cos(E_k(t-t')) J_\varphi(t')
 +
 i\int J_\varphi G_{\rm ret} J_\chi
 \Bigg]
 \non
 \ee
 The propagators are then given by
 \be
 G_{\rm Q}^{\varphi\varphi} & = & (1/2 + n_{\rm BE}(E_p)) 
 2\pi\,\delta(p^2 - m^2) 
 \label{eq:prop_1}
 \\
 G_{\rm Q}^{\varphi\chi} & = &  {i\over p^2 - m^2 + i\epsilon p_0}
 \label{eq:prop_2}
 \\
 G_{\rm Q}^{\chi\varphi} & = &  {i\over p^2 - m^2 - i\epsilon p_0}
 \\
 G_{\rm Q}^{\chi\chi} & = & 0
 \ee
 where $n_{\rm BE}(E_p) = 1/(e^{E_p/T} - 1)$ is the Bose-Einstein
 distribution.
 The corresponding non-equilibrium propagator is 
 \be
 G_{\rm Q}^{\varphi\varphi} & = & (1/2 + f_{\rm Q}) 
 2\pi\,\delta(p^2 - m^2) 
 \label{eq:prop_5}
 \ee
 where $f_{\rm Q}$ is now the density function to be determined.
 In Ref.\cite{MS}, these quantum form of propagators are used with the
 classical field Feynman rules (i) and (ii) above.
 If $f_Q \gg 1/2$, there is very little difference between this set of
 quantum propagators and the set of classical propagators.
 However, one of the points made in Ref.\cite{MS} was that keeping the 1/2
 term in Eq.(\ref{eq:prop_5}) makes the classical kinetic equation to match
 up with the quantum one up to the next-to-leading order in $f$.
 In view of the fact that this is actually mixing quantum and classical
 descriptions, it deserves a more in-depth study.  We will shortly do so in
 the next section.

 Since we have explicit forms of the quantum and the classical
 propagators now, we can discuss
 the validity of the classical perturbation expansion.
 The differences between the 
 classical and the quantum Feynman rules
 are as follows (excluding the tadpole diagrams): 
 \bitem
  \item[(i)] The quantum $\varphi\varphi$ propagator has $1/2 + f$ 
      whereas the classical one has just $f$.
  \item[(ii)] The quantum interaction includes $\lambda\chi^3\varphi/24$
  term whereas this is missing in the classical case.
 \eitem
 The first of these items indicates that the classical approximation is
 valid in the large $f$ limit, $f \gg 1$ so that the appearance of $1/2$
 does not make a difference.  
\begin{figure}[thb]
\epsfxsize=0.7\tw
\begin{center}
\epsfbox{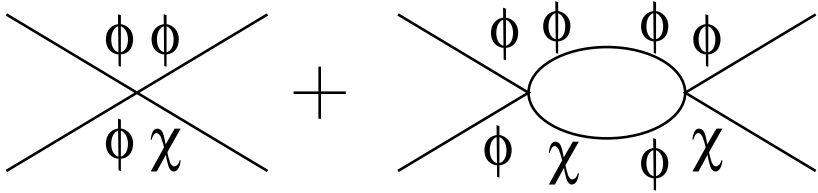}
\end{center}
\caption{The first order vertex correction for the classical field.}
\label{fig:vert_corr}
\end{figure}
 But if $f$ is too big, then potentially perturbative corrections can be as
 large or larger than the leading order contribution.  
 To be concrete, 
 consider the vertex correction depicted in Fig.~\ref{fig:vert_corr}.
 Compared to the bare vertex, the correction term is 
 smaller by a factor of $O(\lambda f_{\rm cl})$.
 Therefore, the {\em classical} perturbation theory is valid when 
 $f_{\rm cl} \gg 1$ but also $\lambda f_{\rm cl} \ll 1$.  This is the same
 conclusion reached in Ref.\cite{MS} but in a heuristic way.
 
\begin{figure}[thb]
\epsfxsize=0.7\tw
\begin{center}
\epsfbox{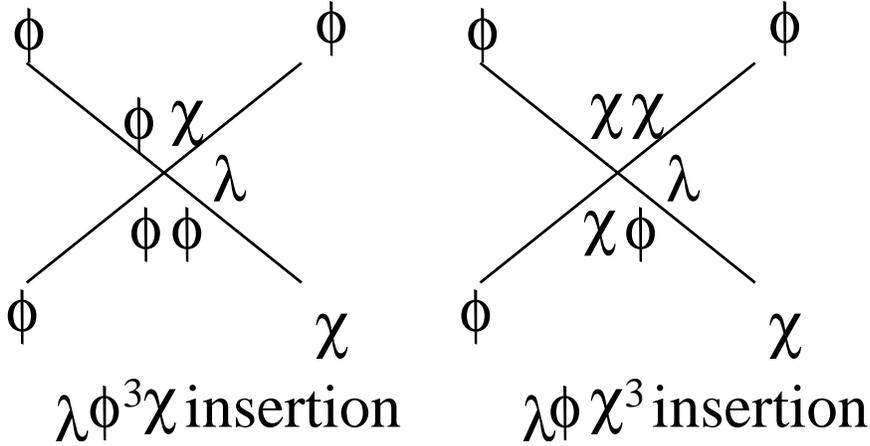}
\end{center}
\caption{Substituting $\lambda\varphi^3\chi$ with $\lambda\chi^3\varphi$.}
\label{fig:insert}
\end{figure}

 The next question we should ask is how big the size of the quantum
 correction is.
 To have an estimate,
 consider substituting a $\lambda\varphi^3\chi$ vertex with a 
 $\lambda\chi^3\varphi$
 vertex in a classical perturbation theory diagram.
 Since there is no $\chi\chi$ propagator, the only way this substitution is
 possible is when the inserted vertices are connected with three
 $\varphi$ field and one $\chi$ field as depicted in Fig.~\ref{fig:insert}.
 In the large $f$ limit, the quantum contribution is $O(1/f^2)$ smaller than 
 the purely classical contribution. 
 Therefore at a given order in $\lambda$, the quantum correction is always
 $O(1/f^2)$ smaller than the classical one.  Again, this is the same
 conclusion as in Ref.\cite{MS}. 
 
 These estimates also indicate that there is a limit that we can trust the
 classical perturbation theory depending on the relative size of $1/f$ and
 $\lambda$.  Suppose that $f = O(1/\sqrt{\lambda})$.  In the small $\lambda$
 limit, this fulfills $f \gg 1$ as well as $\lambda f \ll 1$. 
 However, since $1/f^2 = O(\lambda)$, only the leading order classical 
 perturbative correction is reliable.  The leading order (in $\lambda$)
 quantum correction in
 this case is as big as the second order classical correction.

 \section{Comparison of Quantum and Classical Boltzmann equations}

 Having derived the Feynman rules,
 the derivation of Boltzmann equation can now proceed as in Ref.\cite{MS} for
 the $\lambda\phi^4$ theory.
 For convenience, here we show the results from Ref.\cite{MS}.
 The collision term of the Boltzmann equation is given by
 \be
 C 
 & = &
 {\lambda^2\over 2} \int {d^3k_1 d^3 k_2 d^3 k_3\over (2\pi)^5 2^4
 \omega(k_1)\omega(k_2) \omega(k_3)\omega(p)}
 \delta(\omega(p)+\omega(k_1)-\omega(k_2)-\omega(k_3))
 \delta(\bfp + \bfk_1 - \bfk_2 - \bfk_3)
 \non
 & & {}
 \Bigg\{
 F(k_1)
 F(k_2)
 F(k_3)
 +
 F(p)
 F(k_2)
 F(k_3)
 -
 F(p)
 F(k_1)
 F(k_2)
 -
 F(p)
 F(k_1)
 F(k_3)
 \Bigg\}
 \label{eq:coll}
 \ee
 where we used $F(p) = f(p) + 1/2$ to simplify and combine the terms in
 Ref.\cite{MS}.

 As shown in Section~\ref{sec:props}, quantum and classical propagators have
 different forms. 
 The above expression (\ref{eq:coll})
 uses the quantum form of propagators but otherwise ignores quantum
 corrections. 
 If we are in a purely classical regime, $F$ 
 in the above expression must be changed to $f$.
 The resulting Boltzmann equation then contains only the terms cubic in $f$.
 This is appropriate since it is easy to verify that the equilibrium
 distribution resulting from having only the cubic terms is 
 $f(E) = T/E$. This form leads to the Rayleigh-Jeans catastrophe
 which Planck cured by inventing Quantum Mechanics.

 On the other hand, if we keep the above form, the steady state solution is
 \be
 f(E) = {T\over E} - {1\over 2}
 \ee
 which may be recognized as the first 2 terms of the expansion of the
 Bose-Einstein distribution in the large $T$ limit.
 Therefore in the $T \gg E$ limit, this form is a better
 approximation of the true quantum distribution than the Rayleigh-Jeans form.
 However unlike the Rayleigh-Jeans form, 
 this form of the distribution function becomes negative when $T < 2E$ which
 is clearly unphysical.  Therefore mixing the quantum propagators and the
 classical vertices does not necessarily
 improve the purely classical result.
 
 Let us now consider why keeping the 1/2 term in the propagator somewhat
 mysteriously reproduce the quantum statistical 
 result up to the next-to-leading order
 terms in $f$.
 Suppose we have a scattering
 process that involves $m$ initial particles ($k_i$)
 and $n$ final particles ($p_j$). 
 The quantum Boltzmann equation must contain the following combination of
 the density functions:
 \be
 C_Q 
 & = &
 \prod_{i=1}^m f_Q(k_i) \prod_{j=1}^n [1+f_Q(p_j)]
 -
 \prod_{j=1}^n f_Q(p_j) \prod_{i=1}^m [1+f_Q(k_i)] 
 \ee
 to be consistent with the equilibrium Bose-Einstein density.
 Expanding in terms of $f_Q$'s (we are in the large $f_Q$ limit), we have
 \be
 C_Q
 & = &
 \sum_{s=1}^n
 \prod_{i=1}^m f_Q(k_i) \prod_{j=1,j\ne s}^n f_Q(p_j)
 +
 \sum_{s=1}^n\sum_{t=1}^{n}\theta(s>t)
 \prod_{i=1}^m f_Q(k_i) \prod_{j=1, j\ne s, j\ne t}^n f_Q(p_j)
 \non
 & & 
 -
 \sum_{s=1}^m 
 \prod_{i=1,i\ne s}^m f_Q(k_i) 
 \prod_{j=1}^n f_Q(p_j) 
 -
 \sum_{s=1}^m\sum_{t=1}^{m}\theta(s>t) 
 \prod_{i=1,i\ne s,i\ne t}^m f_Q(k_i) 
 \prod_{j=1}^n f_Q(p_j) 
 \non
 & & {}
 + o(f^{m+n-3})
 \label{eq:CQ}
 \ee
 where we defined $\theta({\rm condition}) = 1$ if the condition is met and
 $\theta({\rm condition}) = 0$ otherwise.
 On the other hand, the classical counter-part must contain
 \be
 C_{\rm cl}
 & = &
 \sum_{s=1}^n
 \prod_{i=1}^m f_{\rm cl}(k_i) \prod_{j=1,j\ne s}^n f_{\rm cl}(p_j)
 -
 \sum_{s=1}^m 
 \prod_{i=1,i\ne s}^m f_{\rm cl}(k_i) 
 \prod_{j=1}^n f_{\rm cl}(p_j) 
 \ee
 to be consistent with the Rayleigh-Jeans form of the equilibrium
 distribution.
 It is clear that the  leading order terms match between the
 quantum and the classical cases.
 Now let $f_{\rm cl} \to f_{\rm cl}+1/2$ to get
 \be
 C'_{\rm cl}
 & = &
 \sum_{s=1}^n
 \prod_{i=1}^m [f_{\rm cl}(k_i)+1/2] 
 \prod_{j=1,j\ne s}^n [f_{\rm cl}(p_j)+1/2]
 -
 \sum_{s=1}^m 
 \prod_{i=1,i\ne s}^m [f_{\rm cl}(k_i) +1/2]
 \prod_{j=1}^n [f_{\rm cl}(p_j) +1/2]
 \non
 \ee
 Expanding up to the next order in $f_{\rm cl}$ yields
 \be
 C'_{\rm cl}
 & = &
 \sum_{s=1}^n
 \prod_{i=1}^m f_{\rm cl}(k_i) 
 \prod_{j=1,j\ne s}^n f_{\rm cl}(p_j)
 -
 \sum_{s=1}^m 
 \prod_{i=1,i\ne s}^m f_{\rm cl}(k_i)
 \prod_{j=1}^n f_{\rm cl}(p_j)
 \non
 & & {}
 +
 {1\over 2}
 \sum_{s=1}^n\sum_{t=1}^n\theta(s\ne t)
 \prod_{i=1}^m f_{\rm cl}(k_i)
 \prod_{j=1,j\ne s,j\ne t}^n f_{\rm cl}(p_j)
 -
 {1\over 2}
 \sum_{s=1}^m\sum_{t=1}^m\theta(s\ne t)
 \prod_{i=1,i\ne s,i\ne t}^m f_{\rm cl}(k_i) 
 \prod_{j=1}^n f_{\rm cl}(p_j) 
 \non
 & & {}
 + o(f^{m+n-3})
 \ee
 which can be recognized to be
 the same as Eq.(\ref{eq:CQ}) once we use the fact that
 $\theta(s>t)$ is equivalent to 
 $\theta(s\ne t)/2$ under the sums. 
 Thus the {\em prescription}, $f_{\rm cl}\to f_{\rm cl}+1/2$ does produce
 the first two terms in the quantum Boltzmann equation for any general
 scattering process.  This also explains why the same kind of consideration
 also seems to work for hot QCD~\cite{Stockamp:2004qu}
 Although it is curious that this prescription can indeed get the
 next-to-leading order term in $f$ right,
 this is irrelevant in getting the right equilibrium limit
 since that requires retaining all the terms in $f$.

\section{Separation of Hard and Soft Modes}

 In view of the discussion in the last section,
 the better application of the statistical mechanics of the 
 classical field is not to get the final equilibrium 
 distribution\footnote{%
     This was also recognized in Ref.{\protect\cite{MS}} where it was
     implied that their formalism is for the system evolving toward
     equilibrium but not applicable near the kinetic equilibrium.} 
 but to use
 it as an intermediate stage where the field degrees of freedom
 and the particle degrees of freedom intermix.
 In this section, we illustrate how this may be achieved. 
 A more detailed analysis of this important problem, however, is out of the
 scope of the present manuscript and will be reported in later
 publications\cite{stefan}.

 As shown in the last section, the quantum perturbative correction becomes
 the same size as the classical perturbative correction when $f\sim 1$.
 As indicated by the equilibrium form of $f$, $f(k)$ becomes small as
 $k$ becomes larger.
 Therefore, the classical field description is usually appropriate only for
 the soft modes.

 With the price of introducing a cutoff $\Lambda$ (where $f$ becomes $\sim 1$),
 the quantum
 field should then be split into the soft and hard modes:
 \be
 \varphi = \varphi_s + \varphi_h
 \ \ \ \ \hbox{and}\ \ \ \ 
 \chi = \chi_s + \chi_h
 \ee
 where the soft modes (with the subscript $s$) have $|\bfk| < \Lambda$
 and the hard modes (with the subscript $h$) have $|\bfk| \ge \Lambda$.  
 The soft modes should then be treated as a classical field and the hard modes
 should be kept as a quantum field. 
 
 To be concrete, consider the $\lambda\phi^4$ theory with the Hamiltonian
 given by
 \be
 H = 
 \int d^3 x
 \left( {\pi^2\over 2} + {(\nabla \phi)^2\over 2} + 
 {m^2\over 2}\phi^2 + 
 {\lambda\over 4!}\phi^4
 \right)
 \ee
 The continuum version of Eq.(\ref{eq:rho_W_fin}) is
 \be
 \rho_W[\varphi_{N+1}, \pi^\varphi_{N+1}]
 & = & 
 \int [d\varphi_0][d\chi_0]\,
 \rho_W[\varphi_0, \pi^\varphi_0]
 \non && {} \times
 \int \cal D\varphi\, 
 \int \cal D\chi\, 
 \int \cal D\pi^\varphi\, 
 \int \cal D\pi^\chi\, 
 \non & & {} \times
 \exp\left[
 i\int \pi^\chi \left( \dot{\varphi} - \pi^\varphi \right)
 -i \int
 \left\{
 \chi 
 \left(\dot{\pi^\varphi} - \nabla^2\varphi + m^2 \varphi + {\lambda\over 3!}
 \varphi^3\right)
 +
 {\lambda\over 4!} \chi^3 \varphi
 \right\}
 \right]
 \non
 \ee

 Separating the hard and the soft modes, the interaction term becomes
 \be
 V 
 &= &
 {\lambda\over 3!}(\chi_h + \chi_s)(\varphi_h + \varphi_s)^3
 +
 {\lambda\over 4!}(\varphi_h + \varphi_s) (\chi_h + \chi_s)^3
 \ee
Keeping only the terms that contain the most number of $\varphi_s$
yields 
\be
V
& = &
{\lambda\over 4!}
\Bigg(
4 \chi_s\varphi_s^3 
+
4 \chi_h\varphi_s^3 
\Bigg)
\ee
These terms are linear in $\chi_s$ and 
$\chi_h$.
Hence, the equations of motion are
\be
& 
\displaystyle
\partial^2 \varphi_s - m^2 \varphi_s - {\lambda\over 3!} \varphi_s^3 = 0 &
\\
\inline{and}
&\displaystyle
\partial^2 \varphi_h - m^2 \varphi_h 
= {\lambda\over 3!} \varphi_s^3 &
\ee
These equations describe a system where the classical field
evolves independent of the particles but the {\em quantum mechanical}
free particles are produces by a classical source.
From the analogous problem in QED (generation of photons from a classical
source), it is an easy matter to show that the spectrum of hard modes at the
end of evolution is
\be
f_h(E_k, \bfk_h) = {\lambda^2 \over 36} 
\left| J(E_k,\bfk_h) \right|^2
\ee
where
\be
J(E_k, \bfk_h) =
\int d^3 x \int_{t_0}^\infty dt\, e^{-i\bfk_h\cdot\bfx + iE_k t}
(\varphi_s(t, x))^3
\ee
This is, of course, in addition to the hard modes which already existed in
the initial state.

If the cut-off $\Lambda$ remains constant at all times, this is the complete
solution for the hard spectrum provided that we can solve the soft mode
classical equation of motion either perturbatively or non-perturbatively.  
In reality, it is not as simple 
due to the fact that the cut-off
$\Lambda$ should be a functional of $\varphi_s$, too.
In the Color Glass Condensate approach to the heavy ion collisions
\cite{Krasnitz:2000gz},
a similar conversion of field degrees of freedom to particle
degrees of freedom was performed to get partons out of classical non-Abelian
gauge field.  The line of argument given here may provide a firmer ground
for such a treatment.
One can conceive that a kind of `renormalization group' 
equation in the manner of Ref.\cite{Boyanovsky:1998aa}
should exist for the cut-off.
This will then enable us to describe the whole system in a consistent
fashion.  This topic is currently under investigation.

\section{Summary and Discussion}

 In this paper, we showed that one can indeed go from the
 statistical mechanics of quantum field theory to the statistical mechanics of
 particles via classical field theory. Along the way, we encountered several 
 subtleties that had to be carefully dealt with. In this regard, we improve
 upon the treatment given in Ref.\cite{MS}.
 
 Going from quantum field theory to the classical field theory
 involves the Wigner functional which plays the role of initial density
 function.  In other words, the probability of an initial state depends on
 both the value of the field and the value of the first time derivative.
 The form of the initial
 Wigner functional in turn determines the propagators. 
 In addition, our ability to use the Feynman diagrams with energy-momentum
 conservation at each vertex depends very much on the fact that time
 translation invariance is maintained. 
 For this to work at least approximately, the Wigner transformed propagators
 $G_W(p, X)$ ({\it c.f.}~Eq.(\ref{eq:G_Wigner}))
 cannot depend strongly on $X = (x+y)/2$.
 This in turn implies that the density matrix at time $t$ must also be
 approximately of the form
 \be
 \ln \rho \sim  E_k^2 \varphi_k^2 + \pi_k^2 \sim H_k
 \ee
 where $\varphi_k$ and $\pi_k$ are Fourier components of the 
 corresponding fields at $t$ and $H_k$ is the corresponding free field
 Hamiltonian. This is a rather strong condition
 which is in general valid only near equilibrium.  Whether this can be
 reconciled with the classical field limit is not yet clear and requires
 further study.

 Going from the quantum field theory to the classical field theory also
 involves a Jacobian in the form of the Van Vleck determinant.
 The role of this Jacobian turned out to be to cancel the single remaining
 quantum effect still left in the classical perturbation theory in the form
 of the tadpole self-energy.

 It must be also emphasized that to define propagators, one must
 define what is meant by the `un-perturbed vacuum'.  Pure classical vacuum 
 cannot fluctuate as it is literally a state where nothing exists.
 On the other hand, the quantum vacuum fluctuates all the time.  
 The importance of this distinction is particularly apparent when one
 considers the free field theory.  The form of the path integral for
 the classical and the quantum theory in this case
 is exactly the same except the form of
 the initial density functional.  To be consistent, the 
 classical initial density
 functional must conform to the classical vacuum property and the quantum
 density functional must conform to the quantum vacuum property. 
 Mixing of the two formalism leads to an interesting conclusion that
 the classical field theory can reproduce the results of {\em quantum
 statistical} Boltzmann equation up to the next leading order in the
 density (Ref.\cite{MS}). 
 The correct Boltzmann equation obtained from classical perturbation theory
 of course
 yields the Rayleigh-Jeans form of the equilibrium distribution function as
 it must.

 As for the validity of the classical perturbation theory, in addition to 
 the two conditions already mentioned we also argued that there is a limit
 that a purely classical perturbation theory makes sense.  For instance, we 
 have shown that if $f \sim 1/\sqrt{\lambda}$, the leading order quantum
 correction is as big as the second order classical correction.  And hence,
 going beyond the first order classical perturbation theory
 doesn't make much sense.  Quantum effect must be considered after the
 leading order. 

 We would also like comment on the meaning of `thermalization' here.
 It is well known that an isolated system {\em as a whole} cannot
 thermalize. 
 A simple example is an eigenstate of the total Hamiltonian which, by
 definition, is stationary.
 This is in contrast to the Boltzmann equation where the stationary solution
 is guaranteed to be the equilibrium distribution.
 The crucial difference of course is whether one is interested in the system
 as a whole, or just a {\em part} of the system be it in the
 momentum space or the coordinate space.  It is therefore perhaps more
 natural to apply the current formalism to a particular sector of the
 system, say the soft modes, and regard the hard spectrum as
 particle degrees of freedom\footnote{This is different from similar
 separation of hard and soft modes previously considered 
 in the {\em equilibrium setting}. 
 For instance, see Ref.\cite{Gleiser:1993ea}.}.

 In summary, in this paper, we re-examined the derivation of the
 Boltzmann equation from the classical field theory as advocated by Mueller
 and Son \cite{MS}.  We pointed out a few subtle points 
 in manipulating functional integrals and developing the perturbation theory
 and showed how to deal with them.
 With our improvement, 
 the framework advocated by Mueller and Son
 has a potential to be a very useful tool in investigating dense many-body
 systems, for instance, the problem of converting field degrees of freedom
 to particle degrees of freedom. 
 Further study in this line is continuing and will be reported in
 later publications.

 \acknowledgments{
The author gratefully acknowledges C.~Gale and G.~Moore for many
discussions. He also thanks R.~Venugopalan
for his helpful suggestions in preparing this manuscript.
The author also thanks D.~Kharzeev and G.~Aarts for their suggestions.
S.J.~is supported in part by the Natural Sciences and
Engineering Research Council of Canada and by le Fonds 
Nature et Technologies of Qu\'ebec.  
S.J.~also
thanks RIKEN BNL Center and U.S. Department of Energy [DE-AC02-98CH10886] for
providing facilities essential for the completion of this work.
}

\end{document}